\documentclass[11pt,a4paper]{article}
\pdfoutput=1

\usepackage{jheppub}
\usepackage{amsfonts}
\usepackage{amssymb}
\usepackage{comment}
\usepackage{epsfig}
\usepackage{graphicx}
\usepackage{amsmath}
\usepackage{tabu}
\usepackage[utf8]{inputenc}
\usepackage{latexsym}
\usepackage{tabularx}
\usepackage[nottoc]{tocbibind}
\usepackage{subfigure}
\usepackage{multirow}
\usepackage{booktabs}
\usepackage{enumerate}
\usepackage{mathrsfs}
\usepackage[utf8]{inputenc}
\usepackage{latexsym} 
\usepackage{amssymb} 
\usepackage{amsmath}

\usepackage {graphicx}
\usepackage {epsfig}
\usepackage {subfigure}
\usepackage{epstopdf}
\usepackage {tabularx} 
\usepackage[nottoc]{tocbibind} 
\usepackage[section]{placeins}
\usepackage{placeins}
\usepackage{color}

\newcommand{\bea}{\begin{eqnarray}}
\newcommand{\eea}{\end{eqnarray}}
\newcommand{\bi}{\begin{itemize}}
\newcommand{\ei}{\end{itemize}}
\newcommand{\ben}{\begin{enumerate}}
\newcommand{\een}{\end{enumerate}}
\newcommand{\be}{\begin{equation}}
\newcommand{\ee}{\end{equation}}
\newcommand{\ba}{\begin{align}}
\newcommand{\ea}{\end{align}}
\newcommand{\comments}[1]{}

\newcommand{\mc}{\mathcal}

\newcommand{\beqa}{\begin{eqnarray}}
\newcommand{\eeqa}{\end{eqnarray}}
\newcommand{\V}{{\cal{V}}}

\title{Light Higgsino Dark Matter from Non-thermal Cosmology}

\author[1]{\small{Luis Aparicio,}}
\author[1,2,3]{\small{Michele Cicoli,}}
\author[4]{\small{Bhaskar Dutta,}}
\author[2,3]{\small{Francesco Muia,}}
\author[1,5]{\small{Fernando Quevedo}}

\affiliation[1]{ICTP, Strada Costiera 11, Trieste 34014, Italy}
\affiliation[2]{Dipartimento di Fisica e Astronomia, Universit\`a di Bologna, \\ via Irnerio 46, 40126 Bologna, Italy}
\affiliation[3]{INFN, Sezione di Bologna, via Irnerio 46, 40126 Bologna, Italy}
\affiliation[4]{Department of Physics and Astronomy, Mitchell Institute for Fundamental Physics \\
and Astronomy, TAMU, College Station, TX 77843-4242, USA}
\affiliation[5]{DAMTP, Centre for Mathematical Sciences, Wilberforce Road, Cambridge, CB3 0WA, UK.}

\emailAdd{laparici@ictp.it}
\emailAdd{mcicoli@ictp.it}
\emailAdd{dutta@physics.tamu.edu}
\emailAdd{muia@bo.infn.it}
\emailAdd{f.quevedo@damtp.cam.ac.uk}

\abstract{We study the scenario of higgsino dark matter in the context of a non-standard cosmology with a period of matter domination prior to Big Bang nucleosynthesis. Matter domination changes the dark matter relic abundance if it ends via reheating to a temperature below the higgsino thermal freeze-out temperature. We perform a model independent analysis of the higgsino dark matter production in such scenario. We show that light higgsino-type dark matter is possible for reheating temperatures close to 1 GeV. We study the impact of dark matter indirect detection and collider physics in this context. We show that Fermi-LAT data rule out non-thermal higgsinos with masses below 300 GeV. Future indirect dark matter searches from Fermi-LAT and CTA will be able to cover essentially the full parameter space. Contrary to the thermal case, collider signals from a 100 TeV collider could fully test the non-thermal higgsino scenario. In the second part of the paper we discuss the motivation of such non-thermal cosmology from the perspective of string theory with late-time decaying moduli for both KKLT and LVS moduli stabilisation mechanisms. We finally describe the impact of embedding higgsino dark matter in these scenarios.}

\preprint{DAMTP-2015-50, MI-TH-1535 \\
\phantom{a} \hfill{}}


\begin{document}

\maketitle

\bigskip

\section{Motivation and summary}
\label{section1}

The best candidate for dark matter (DM) in supersymmetric models with R-parity conservation is the lightest neutralino $\chi_1^0$, which is generically the lightest supersymmetric particle (LSP). Neutralinos are weakly interacting particles (WIMPs) which, in the standard thermal picture, are assumed to be in equilibrium with the thermal bath in a radiation dominated universe. As the universe expands, it cools down and at some point the temperature drops below the WIMP mass $m_\chi$. At that moment neutralinos become non-relativistic and their abundance per comoving volume decreases due to the Boltzmann factor $\exp\left(-m_\chi/T\right)$ until it reaches its freeze-out value at the temperature $T_{\rm f}$ which is typically of order $T_{\rm f} \simeq m_\chi/20$. This happens when the WIMP annihilation rate becomes of order the Hubble parameter $H$ and DM particles drop out of thermal equilibrium. 

Hence the thermally produced DM abundance depends just on its thermal averaged annihilation rate $\langle \sigma_{\rm ann} v \rangle$:
\be
\Omega = \Omega^{\rm obs} \,\frac{\langle \sigma_{\rm ann} v\rangle^{\rm th}}{\langle \sigma_{\rm ann} v \rangle} \,,
\label{eq:ThermalAbundance}
\ee
where $\Omega^{\rm obs} \simeq 0.23$ is the abundance observed by the Planck satellite \cite{Ade:2015xua}, while $\langle \sigma_{\rm ann} v\rangle^{\rm th} = 3 \times 10^{-26} \, \text{cm}^3 \,\text{sec}^{-1}$ is the reference value which gives the correct relic abundance. This makes the thermal scenario very predictive and completely independent of the previous thermal history of the universe.

From \eqref{eq:ThermalAbundance} we can see that:
\be
\Omega \simeq 0.23\ \frac{\alpha^2/(200 \, \rm{ GeV})^2}{\langle \sigma_{\rm ann} v \rangle} \,,
\ee
where $\alpha = g_2^2/(4\pi)$. Given that $\langle \sigma_{\rm ann} v \rangle \simeq \alpha^2/m_\chi^2$, weakly interacting particles with masses around the weak scale $m_\chi\sim m_{\rm weak}\sim\mc{O}(100)$ GeV naturally give rise to the observed DM relic density. This fact is very well known in the literature under the name of ‘WIMP miracle’ and it suggests that new degrees of freedom at the weak scale are natural DM candidates.

However, in the context of supersymmetry (SUSY), WIMP candidates do not really satisfy the condition $m_\chi\sim m_{\rm weak}$: thermal higgsinos saturate the DM relic density for masses around $1$ TeV, while winos need to be around $2.5$ - $3$ TeV. The situation for binos is even worse because their annihilation cross section is so small that they always overproduce DM.\footnote{That is also the case of singlinos in SUSY models with an extra scalar like the NMSSM.} This problem can be avoided either by focusing on fine-tuned corners of the underlying parameter space, like A-funnels or coannihilation with other sparticles, or by considering so-called well tempered combinations of electroweakinos which can lead to the correct DM abundance. However, recent direct detection results show that these scenarios are either under siege or directly ruled out. Thus a correct thermal production of the observed DM abundance seems to require a high level of fine-tuning. 

In the present paper we shall therefore consider a different production mechanism based on a non-standard cosmological evolution of our universe. More precisely, we shall consider the situation where DM particles are produced via a non-thermal mechanism based on the late time decay of heavy scalars with only gravitational couplings to ordinary matter. This production mechanism is well motivated from both a bottom-up and a top-down perspective. Since current observations can trace back the thermal history of the universe only up to Big Bang Nucleosynthesis (BBN), when the temperature of the thermal bath was around $T_{\rm BBN} \simeq 3$ MeV, there is no reason in principle to assume a standard cosmological evolution for temperatures above $T_{\rm BBN}$. In particular, the generic presence of gravitationally coupled particles (like moduli or gravitinos) in UV complete theories like string theory, can change the cosmological evolution of our universe.

Moduli are scalar fields that get displaced from their late-time minimum during inflation due to the inflationary energy density \cite{Dine:1995kz}. After the end of inflation, their VEV decreases following the Hubble parameter $H$ until $H$ becomes of order their mass and the moduli start oscillating around their late-time minimum. Since their energy density redshifts as matter, they quickly come to dominate the energy density of the universe, introducing a new era of matter domination before BBN. Finally, these moduli decay when $H$ becomes of order their decay rate $\Gamma_\phi \simeq m_\phi^3/M_p^2$ with $M_p=2.4 \times 10^{18}$ GeV. The decay of the moduli heats the thermal bath and produces entropy diluting everything that has been produced before. Moreover, the moduli decay leads also to the non-thermal production of the lightest neutralino. 

This scenario gives rise to an interesting cosmological evolution of the universe which has been vastly studied in the literature \cite{McDonald:1989jd,Chung,Moroi:1999zb,Giudice:2000ex,kamionkowski-turner,Moroi,Drees,Khalil,Fornengo:2002db,Pallis:2004yy,endo,kohri,Gelmini:2006pw,Gelmini:2006pq,dutta,Acharya:2009zt,Arcadi:2011ev,Fan:2013faa,Blinov:2014nla,Kane:2015jia}. In the non-thermal scenario, differently from the thermal case, the DM relic density depends on two parameters: the WIMP annihilation rate and the reheating temperature (or equivalently the moduli mass). This additional parameter gives enough freedom to reproduce the observed DM relic density for neutralino masses of order $m_\chi \simeq m_{\rm weak}$. Given that non-thermally produced WIMPs can be light, this scenario turns out to be interesting for DM indirect detection and collider physics bounds. We will show that this new ‘WIMP miracle’ can happen only if the moduli masses are around $10^6$ - $10^7$ GeV. The ‘naturalness’ of this energy scale for the moduli masses depends on moduli stabilisation (and therefore ultimately on the string landscape).

The first part of the paper is a model independent analysis of non-thermal higgsino DM,\footnote{Higgsinos are good DM candidates in models like split-SUSY where there is a hierarchy between electroweakinos and scalars \cite{Wells:2003tf}. Moreover, light higgsinos are well motivated in natural SUSY scenarios \cite{Hall:2011aa}.} leaving the wino and bino DM cases for future work. The main conclusions of this model independent analysis are the following:
\begin{enumerate}
\item The observed DM relic density can be saturated even for higgsino masses as low as $100$ GeV.

\item The strongest lower bound on the mass of non-thermal higgsinos comes from indirect detection which requires $m_\chi \gtrsim 300$. This bound comes from the non-observation by Fermi-LAT \cite{Ackermann:2015zua} of gamma rays due to dark matter annihilation from dwarf spheroidal galaxies where the dependence on the astrophysical profile is less important than in galactic centre observations.

\item We also show that future observations from Fermi-LAT or CTA \cite{Carr:2015hta} could cover essentially the entire parameter space of this scenario. Moreover, unlike the thermal case, collider signals from the LHC can probe only a small part of the parameter space using monojet plus soft lepton searches \cite{Baer:2014kya}. On the other hand, a $100$ TeV machine could test directly all the parameter space using monojet and disappearing tracks searches \cite{Low:2014cba}.
\end{enumerate}

In the second part of the paper we go into a model dependent discussion. We study the non-thermal post-inflationary cosmological evolution of two well-established scenarios of string moduli stabilisation: KKLT \cite{Kachru:2003aw} and the Large Volume Scenario (LVS) \cite{Balasubramanian:2005zx}. In both cases, we determine the mass hierarchy between moduli, higgsinos and other sparticles. The main difference between these two scenarios is that in LVS the late decaying particle is the lightest modulus while in KKTL it is the gravitino. At the level of non-thermal DM production this does not change anything but it has consequences on setting the gaugino and SUSY-breaking scale. In each case, we have also worked out the consequences of preserving the BBN results in the presence of late decaying particles \cite{CMP}. The main conclusions can be summarised as follows:
\begin{enumerate}
\item If the visible sector is localised on D7-branes, both cases lead to non-thermal DM overproduction, and so R-parity violation is mandatory.

\item If the visible sector is localised on D3-branes, both KKLT and LVS models can give rise to an allowed region of the parameter space where non-thermally produced light higgsinos can correctly reproduce the observed DM abundance. 

\item LVS models with the visible sector on D3-branes are particularly interesting since the hierarchy between the lightest modulus and the SUSY particles allows to set bounds from DM direct detection which however depend on the moduli VEVs (and so they are less constraining than the ones from indirect detection and collider searches). We have performed an analysis for a particularly well motivated value of the volume of the extra-dimensions and the result is twofold: $(i)$ in order to obtain constraints which are stronger than the ones from indirect detection, one would need data from large scale DM direct detection experiments (beyond $1$ Ton); $(ii)$ a large portion of the parameter space falls below the neutrino background, and so DM direct detection experiments seem to be less useful in this case.
\end{enumerate}

\section{Dark matter in a non-standard cosmology}
\label{section2}

Moduli are scalar fields that couple to all other particles only gravitationally. During inflation, they are displaced from their minimum because of the inflationary energy density. After the end of inflation, once their mass becomes comparable to the Hubble scale ($m_\phi \sim H$), the Hubble friction ceases to be the dominant effect and the moduli start to oscillate around their minimum. After some oscillations the moduli evolution is indistinguishable from pressureless matter and the moduli number per comoving volume remains constant. Hence moduli redshift as matter, with an initial abundance given by:
\be
\rho_\phi = \frac12 \,m_\phi^2\, \phi_0^2\,,
\ee
where $\phi_0$ is the initial misalignment which is in general of order $\phi_0 \sim M_p$ \cite{Dine:1995kz}. After the beginning of the oscillations, the moduli quickly come to dominate the energy density of the universe which therefore becomes matter dominated. When the Hubble parameter becomes of order the moduli decay rate, i.e. $H \sim \Gamma_\phi \simeq m_\phi^3/M_p^2$, these fields decay and a new radiation dominated era begins. This scenario changes the standard cosmological picture because it introduces extra matter dominated epochs between the end of inflation and the BBN epoch. 

The reheating temperature $T_{\rm r}$ of the final radiation dominated era before the BBN epoch is set by the decay $\Gamma_\phi$ of the lightest modulus into Standard Model light degrees of freedom (and possible superpartners): $T_{\rm r} \simeq \sqrt{\Gamma_\phi M_p}$. This non-standard cosmology can potentially modify the DM relic abundance if the WIMP freeze-out temperature $T_{\rm f}$ is larger than the reheating temperature from moduli decay: $T_{\rm f} > T_{\rm r}$. In this case, the freeze-out mechanism takes place during a matter dominated, instead of a standard radiation dominated, era. Moreover, the moduli decay dilutes the neutralino relic density due to entropy production giving \cite{Giudice:2000ex,McDonald:1989jd}: 
\be
\Omega \simeq \frac{T_{\rm f}}{T_{\rm f,new}}\left(\frac{T_{\rm r}}{T_{\rm f,new}}\right)^3 \Omega^{\rm th}\,,
\label{dilut}
\ee
where $\Omega^{\rm th}$ is the standard thermal DM relic density, while $T_{\rm f} \simeq m_\chi/20$ and $T_{\rm f,new}$ is the new freeze-out temperature taking into account the entropy production due to the decay of the lightest modulus. By solving the Boltzmann equations it can however be shown that the difference between $T_{\rm f}$ and $T_{\rm f,new}$ is relevant only for $T_{\rm r} < 1$ GeV \cite{Gelmini:2006pq}. This scenario has been classified under the name of \textit{thermal production without chemical equilibrium} \cite{Gelmini:2006pw}. 

Nevertheless, this dilution is not the only effect produced by the presence of a late time decaying scalar. The direct or indirect decay of moduli into neutralinos yields also a non-thermal production that gives an extra contribution to the neutralino DM abundance. Depending on how efficiently neutralinos annihilate at the time of reheating, i.e. on whether the DM pair annihilation rate $\Gamma_\chi = n_\chi \langle \sigma v \rangle$ is larger or smaller than the expansion rate $H$ at $T_{\rm r}$, DM non-thermal production can follow two scenarios:
\begin{enumerate}
\item If DM particles annihilate very efficiently during the modulus decay, i.e. $\Gamma_\chi > H(T_{\rm r})$, there is a period of chemical equilibrium generated by the combination of the modulus decay and DM annihilation. This period continues until $\Gamma_\chi \sim H$, when DM annihilation is no longer efficient and neutralinos go out of chemical equilibrium (this is usually called \textit{non-thermal freeze-out}). At this point, the neutralino abundance per comoving volume reaches its definitive value. This scenario was first studied in \cite{McDonald:1989jd,Moroi:1999zb,Giudice:2000ex} and received several names: \textit{non-thermal production with chemical equilibrium} \cite{Gelmini:2006pw}, \textit{annihilation scenario} \cite{Kane:2015jia} or \textit{re-annihilation scenario} \cite{Cheung:2010gj}.
 
\item If DM particles produced from the modulus decay do not interact further, i.e. $\Gamma_\chi < H(T_{\rm r})$, their abundance is just the one produced by the modulus decay. Since there is no efficient annihilation, the DM number density per comoving volume is frozen from the beginning. This scenario has been known both as \textit{non-thermal production without chemical equilibrium} \cite{Gelmini:2006pw} and as \textit{branching scenario} \cite{Kane:2015jia}.
\end{enumerate}

The DM abundance per comoving volume for both scenarios can be expressed as \cite{Moroi:1999zb}:
\be
\left(\frac{n_\chi}{s}\right) = \text{min} \left[\left(\frac{n_\chi}{s}\right)^{\rm obs} \, \frac{\langle \sigma_{\rm ann} v\rangle^{\rm th}}{\langle \sigma_{\rm ann} v\rangle} \, \sqrt{\frac{g_*(T_{\rm f})} {g_*(T_{\rm r})}} \, \frac{T_{\rm f}}{T_{\rm r}}, \, Y_\phi \text{Br}_{\chi}\right] \,,
\label{eq:NonThermalAbundance}
\ee
where $\left(\frac{n_\chi}{s}\right)^{\rm obs} \simeq \Omega^{\rm obs} \, \left(\frac{\rho_{\rm crit}}{m_\chi s_0 h^2}\right)$, while $Y_\phi \simeq \frac{3 T_{\rm r}}{4 m_\phi}$ is the yield of particle abundance from modulus decay and Br$_\chi$ is the branching ratio of the modulus decay into DM particles (interpreted as the averaged number of DM particles produced per modulus decay). The annihilation scenario corresponds to the first term in \eqref{eq:NonThermalAbundance}, while the branching scenario is described by the second term of the same expression.

As we have already mentioned, the efficiency of DM annihilation determines whether DM is non-thermally produced in the annihilation or in the branching scenario. In particular, the condition $\Gamma_\chi > H(T_{\rm r})$ can be understood as:
\be
n_\chi (T_{\rm r}) > \frac{H(T_{\rm r})}{\langle \sigma v \rangle}=\frac{\Gamma_\phi}{\langle \sigma v \rangle}\,,
\ee
where $n_\chi (T_{\rm r}) = \text{Br}_{\chi} n_\phi(T_{\rm r}) =  \text{Br}_{\chi} \rho_\phi(T_{\rm r})/m_\phi $. Using the definition of $T_{\rm r}$ and assuming that the modulus thermalises immediately after its decay, this condition can be written as:
\be
\frac{1}{\langle \sigma v \rangle} < \text{Br}_\chi\ \frac{\pi^2}{30}\ T_{\rm r}^{4/3} M_p^{2/3}\,.
\label{cond1}
\ee
For $T_{\rm f}>T_{\rm r}$, the condition to be in the annihilation regime, without any loss of generality, becomes: 
\be
\frac{1}{\langle \sigma v \rangle} < \text{Br}_\chi\ \left(\frac{m_\chi}{ \text{1 GeV}}\right)^{4/3} 10^{10}\ \text{GeV}^2\,.
\label{Eqcond}
\ee
Using the s-wave approximation for the annihilation cross section, we can estimate the regime of masses for which different neutralinos satisfy this condition. From pure winos, higgsinos and binos, it is easy to see that \cite{ArkaniHamed:2006mb}:
\be
m_{\rm wino} < \left( \frac{3 g^4}{16 \pi} \text{Br}_\chi\ 10^{10} \right)^{3/2} \text{GeV} \simeq 10^{12}\ \text{Br}_{\chi}^{3/2}\ \text{GeV}\,,
\label{winos}
\ee
and:
\be 
m_{\rm higgsino} < \left( \frac{g^4}{512 \pi} (21+3\tan^2\theta_W +11\tan^4\theta_W) \text{Br}_{\chi}\ 10^{10} \right)^{3/2} \text{GeV} \simeq 10^{11}\ \text{Br}_\chi^{3/2}\ \text{GeV}\,.
\label{higgsinos}
\ee
Unless the branching ratio $\text{Br}_\chi$ is very small, the conditions \eqref{winos} and \eqref{higgsinos} clearly indicate that winos and higgsinos are always non-thermally produced in the annihilation scenario. The case of binos is instead slightly more model dependent since the condition to be satisfied for being in the annihilation scenario depends on the slepton mass $m_{\tilde{l}_{\rm r}}$: 
\be
m_{\rm bino}>\left(\frac{40\pi}{3g^4 \tan^2\theta_W} \frac{1}{10^{10}\ \text{Br}_\chi} \left(\frac{m_{\tilde{l}_{\rm r}}}{\text{1 GeV}}\right)^4 \right)^{3/10} \text{GeV} \simeq 7\times 10^{-3}\ \text{Br}_\chi^{-3/10} \left(\frac{m_{\tilde{l}_{\rm r}}}{\text{1 GeV}}\right)^{6/5} \, \text{GeV}\,.
\label{binos}
\ee
For $m_{\tilde{l}_{\rm r}}\simeq 100$ GeV (using LEP bounds), the condition to be in the annihilation scenario becomes $m_{\rm bino} > 1.82\ \text{Br}_\chi^{-3/10}$ GeV which basically includes all cases. In order to be in the branching scenario, it is necessary to be have either $\text{Br}_\chi\ll1$ or to go to multi-TeV sleptons. For example for $m_{\tilde{l}_{\rm r}}\simeq 1$ TeV, binos get non-thermally produced in the branching scenario if $m_{\rm bino} < 29\ \text{Br}_\chi^{-3/10}$ GeV, while for $m_{\tilde{l}_{\rm r}}\simeq 10$ TeV, it is necessary to have $m_{\rm bino} < 458\ \text{Br}_\chi^{-3/10}$ GeV.

As a consequence, in the annihilation scenario the DM relic abundance due to moduli decay depends on the neutralino annihilation efficiency at $T_{\rm r}$ and it can be written as \cite{Gelmini:2006pw}:
\be
\Omega^{\rm nt} =\sqrt{\frac{g_*(T_{\rm f})} {g_*(T_{\rm r})}} \,\frac{T_{\rm f}}{T_{\rm r}} \ \Omega^{\rm th}\,,
\label{NTann}
\ee
where $\Omega^{\rm th}$ is the expression for the thermal relic density. On the other hand, for the branching scenario the DM relic density depends on the averaged number of neutralinos per modulus decay but not on the annihilation cross section:
\be
\Omega^{\rm nt} h^2 =1.5\times 10^2\ \text{Br}_\chi\ \left( \frac{T_{\rm r}}{\text{GeV}}\right)^{1/3} \left( \frac{m_\chi}{\text{ GeV}}\right) \kappa^{1/3}\,,
\label{NTann2}
\ee
where $\kappa$ (which is typically of order one) parametrises the model dependence of the modulus decay $\Gamma_\phi =\kappa\ m_\phi^3/M_p^2$.\footnote{Note that $\kappa$ should also appear in (\ref{Eqcond}) but, given that it plays no significant role in that expression, we ignored it for the sake of simplicity.} Note that the DM relic abundance does not overclose the universe only for a very small $\text{Br}_\chi$. In order to avoid fine-tuning issues, such a small number should be justified by a proper theoretical motivation.

\section{Non-thermal higgsino dark matter}
\label{sec:IndirectDetection}

In this section we focus on the analysis of the higgsino LSP case. A mainly higgsino-like neutralino scenario is characterised by a spectrum where the lightest sparticles are the first two neutralinos $\chi_1^0$, $\chi_2^0$ and the first chargino $\chi_1^\pm$. All of them are dominated by the higgsino component and their masses are very close. The degree of degeneracy between the masses depends mainly on the mass of the bino $M_1$ and wino $M_2$: the heavier they are the more degenerate $\chi_1^0$, $\chi_2^0$, $\chi_1^\pm$ will be. 

Let us point out that we shall not consider well-tempered higgsino-gaugino scenarios since they are in strong tension with recent direct detection data. In fact, LUX sets a lower bound on gaugino masses, depending on whether the lightest gaugino is the bino or the wino \cite{lux}. For thermal higgsinos with mass of order $\mu \sim$ 1 TeV, in the higgsino-bino case, $M_1 > 1.2$ TeV in most of the parameter space, while in the higgsino-wino scenario, $M_2>1.6$ GeV \cite{diCortona:2014yua,Badziak:2015qca}. These bounds can be escaped only in a small region with $\mu<0$ and $\tan\beta\leq 2$. However, XENON1T (which will release data probably this year) should be able to probe also this remaining region. Let us finally mention that the LUX bounds on gaugino masses for the higgsino LSP case indicate that the masses of $\chi_1^0$, $\chi_2^0$, $\chi_1^\pm$ are quite close to each other. This has an important impact on collider phenomenology, as we will describe in Sec. \ref{CollPheno}.

Due to the bino/wino bounds mentioned above, from now on we will assume that the LSP is mainly higgsino.\footnote{Note that we are not even assuming MSSM. Even in more complicated SUSY models like NMSSM or beyond, the only assumption is a higgsino LSP.} The rest of the spectrum in this scenario could be either as light as the lightest gauginos like in natural SUSY scenarios \cite{Hall:2011aa}, or it could feature very heavy sfermions like in split-SUSY models \cite{Wells:2003tf}. The second (and last) assumption that we will make is the presence of moduli which can give rise to a non-thermal cosmological history as explained in Sec. \ref{section2}.

\subsection{Indirect detection constraints}

Given that the higgsino LSP case does not need to assume any value of the bino or wino mass (beyond the LUX bounds), the constraints coming from DM direct detection are not very useful. In other words, these bounds are model/spectrum dependent. That is the reason why we are going to focus only on indirect detection constraints and collider signals. In Sec. \ref{sec:DirectDetection} we will discuss some models from UV stringy completions and we will analyse the impact of DM direct detection constraints in terms of the spectrum generated by such stringy scenarios.

\begin{figure}[!t]
\centering
\hspace{-0.5cm}
\includegraphics[width=8.4cm]{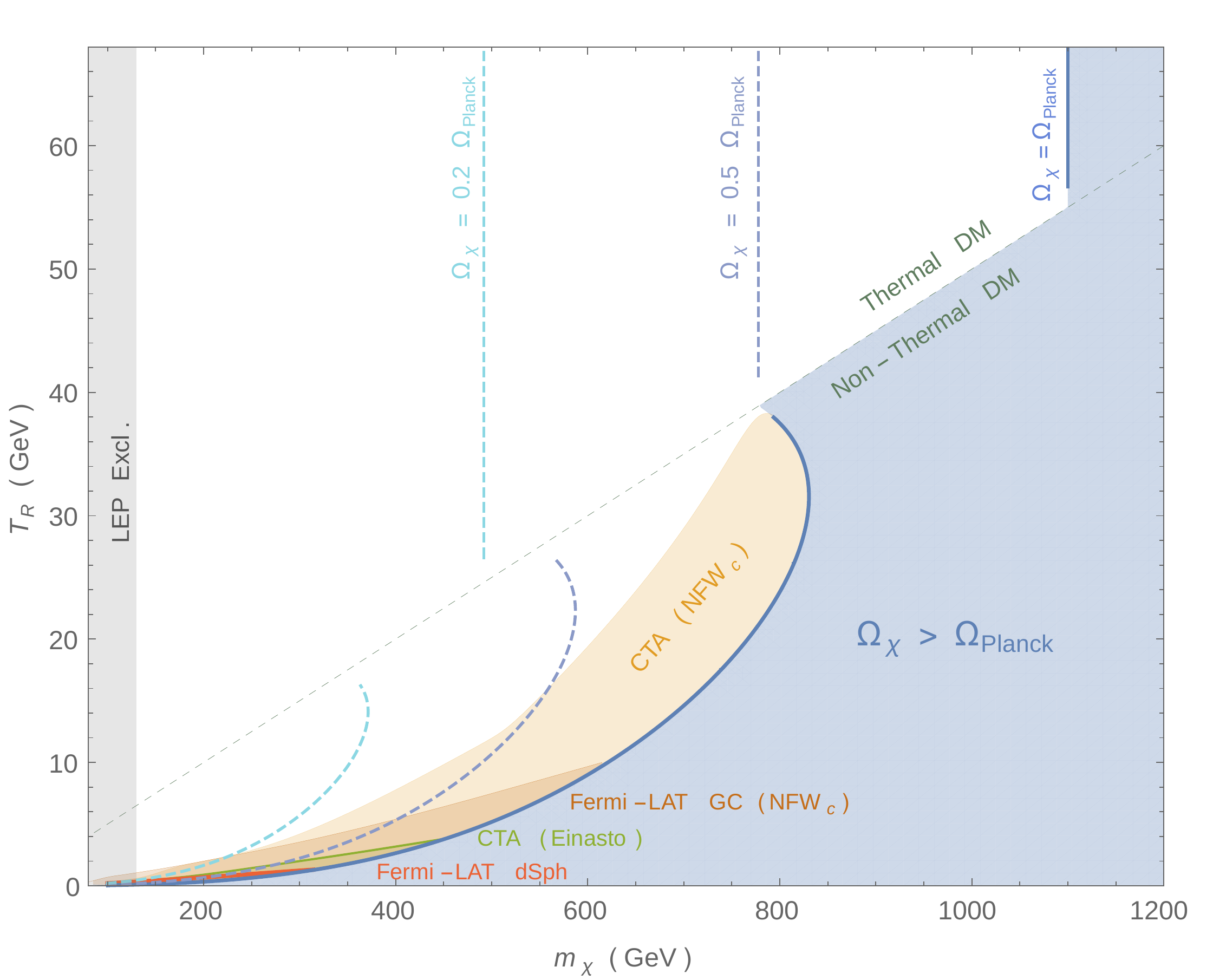}
\hspace{-0.1cm}
\includegraphics[width=8.4cm]{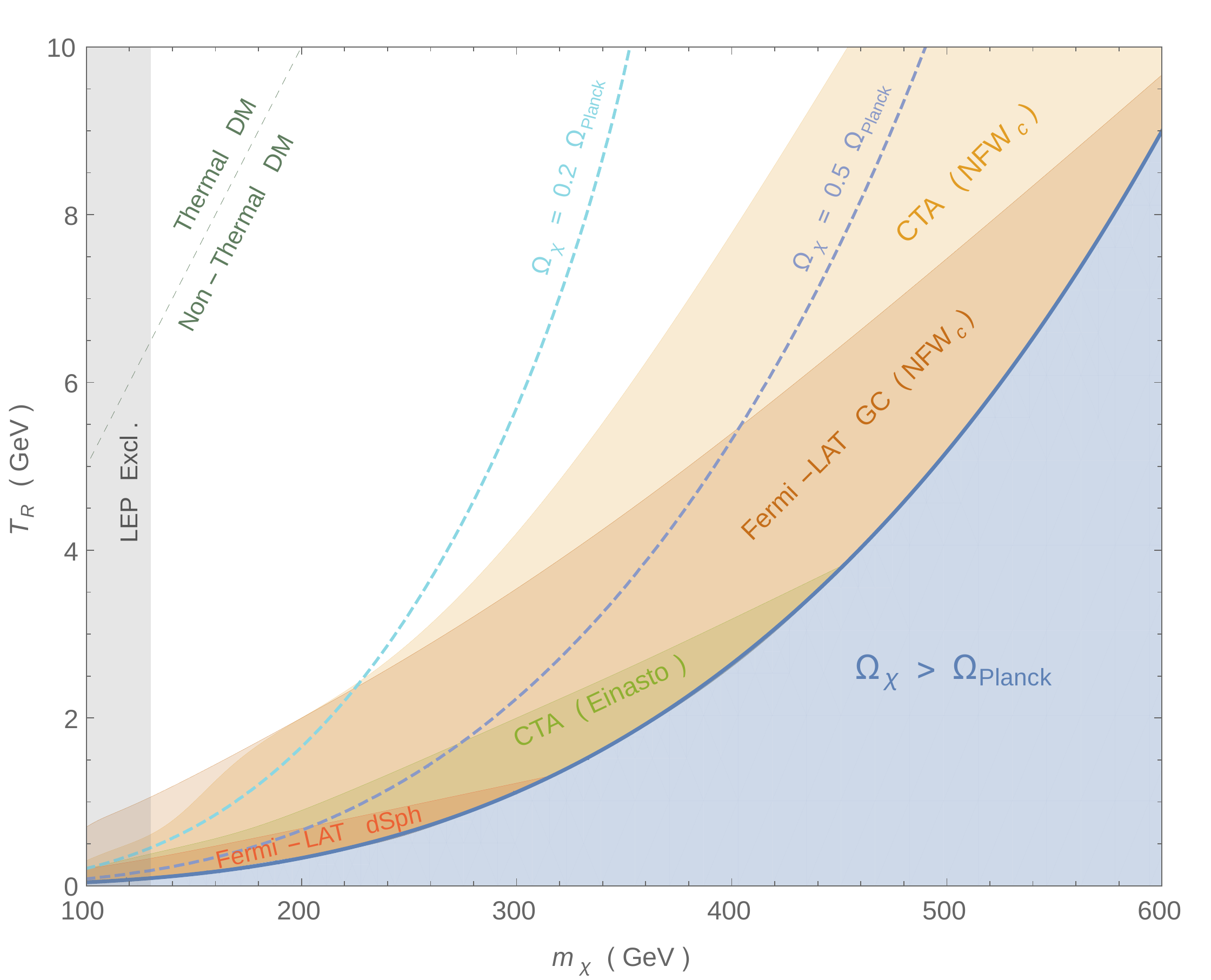}
\hspace{-0.5cm}
\caption{Reheating temperature and DM relic density in terms of the higgsino mass. The diagonal dashed grey line refers to $T_{\rm r} = T_{\rm f} = m_\chi/20$. The coloured regions indicate different bounds from indirect detection experiments. The red and brown regions correspond to the Fermi-LAT pass 8 bounds. The green and orange regions correspond instead to CTA prospects. The plot on the right is a zoom on the region with $T_{\rm r}\leq10$ GeV.}
\label{figInd}
\end{figure}

In Fig. \ref{figInd} we show the results of the analysis of the higgsino LSP scenario in a non-standard cosmology where the lightest modulus decays and reheats the universe at a given $T_{\rm r} \sim \sqrt{m_\phi^3/M_p}$. As mentioned in Sec. \ref{section2}, depending on the relation between $T_{\rm r}$ and the higgsino mass (through $T_{\rm f} \simeq m_\chi/20$), the effect on the DM relic abundance changes. In Fig. \ref{figInd} we show that for $T_{\rm r} > T_{\rm f}$, i.e. for values above the diagonal dashed grey line (which corresponds to $T_{\rm r} = T_{\rm f}  = m_\chi/20$), there is no effect from the presence of moduli. In fact, they would decay before the higgsino thermal freeze-out, and so they would not affect the standard DM thermal production.

However, for $T_{\rm r} < T_{\rm f}$ the modulus decay has a double effect: it dilutes the higgsino relic abundance generated by the thermal freeze-out (the so-called thermal production without chemical equilibrium) and, at the same time, it decays into higgsinos increasing their abundance (the so-called annihilation scenario or non-thermal production with chemical equilibrium). These effects are antagonistic since the former reduces the DM relic density while the second tends to increase it (see \eqref{dilut} and \eqref{NTann}). The combination of these two effects is plotted in Fig. \ref{figInd}. The light blue area of the plot is the region of the parameter space where DM is overproduced, and the blue solid line represents the region where the DM abundance observed by Planck is saturated. The dashed cyan and violet lines represent the regions of the parameter space where higgsino-like DM constitutes only the $50\%$ and $20\%$ of the total DM content. 

If we focus on the solid blue line, it can be seen that, for $T_{\rm r} < T_{\rm f}$, the region with $40$ GeV $\lesssim T_{\rm r} \lesssim 55$ GeV is dominated by the thermal production without chemical equilibrium, i.e. the modulus decay does not heavily dilute the previous thermal higgsino production. As a consequence, in that region of parameter space higgsino DM is overproduced due to the additional DM component coming from the decay of the modulus into higgsinos. Note that the discontinuity of the solid blue line in this region of parameter space has no physical meaning since it is just due to the technical difficulty to consider both non-thermal effects.

At $T_{\rm r} \simeq 38$ GeV (when the modulus mass is $m_\phi \simeq \mc{O} (10^7) $ GeV) the effect of the dilution reduces the thermal relic abundance to half of its initial freeze-out value and, at the same time, the non-thermal production generates precisely the other half required to saturate the DM relic density observed by Planck. From this point on (decreasing $T_{\rm r}$) the effect of the dilution is bigger and bigger, leaving more space for a non-thermal production. In particular, for $T_{\rm r} \simeq 4$ GeV the modulus decay has diluted $80\%$ of the previous thermal DM production, and so most of the DM abundance is due to non-thermal production. From \eqref{dilut} and \eqref{NTann} it is easy to understand that when the annihilation scenario becomes the dominant effect (for lower temperatures), lighter higgsinos are needed to generate the correct DM relic density.

However, there are limits on how light these higgsinos can be. The first is the LEP bound on direct production of charginos, represented in Fig. \ref{figInd} by the grey band, which requires $m_\chi \gtrsim 100$ GeV. Moreover, for light higgsinos which saturate the DM relic density, indirect detection constraints have an important impact. We have analysed this kind of constraints by first computing the thermal averaged cross section of higgsinos with \texttt{micrOMEGAs} \cite{Belanger:2013oya}, and then using the bounds from Fermi-LAT data and the prospects on future experiments like CTA (we have used the limits reported in \cite{Carr:2015hta,Lefranc:2015pza,Silverwood:2014yza}). 

The result is shown in Fig. \ref{figInd}, where we show that the bound coming from Fermi-LAT data on dwarf spheroidal galaxies (Fermi-LAT dSph) sets a lower bound on the higgsino mass of order $m_\chi \gtrsim 300$ GeV \cite{Ackermann:2015zua}. This bound corresponds to $T_{\rm r} \simeq 2$ GeV, which in terms of the modulus mass is $m_\phi \simeq 2\times 10^6$ GeV. Fermi-LAT dSph is the most robust bound given that it does not depend on the DM astrophysical profile and possible astrophysical uncertainties are already taken into account in the limits offered by this collaboration. Fig. \ref{figInd} shows also the Fermi-LAT limit due to the non-observation of DM annihilation from the galactic centre (Fermi-LAT GC) \cite{Gomez-Vargas:2013bea}. This bound ($m_\chi \gtrsim 625$ GeV) is instead very dependent on the actual DM astrophysical profile. In particular, we plot the contracted NFW (NFW$_c$) profile which corresponds to the most cuspy one. Due to the problems on sub-halo galactic structures, cuspy profiles seem to become less motivated \cite{Oh:2015xoa}. Nevertheless, we plot this bound because any other (more cored) profile gives a bound below the Fermi-LAT dSph one. Finally, we also show possible bounds coming from future indirect detection experiments like CTA, which again correspond to cuspy DM astrophysical profiles because the cored ones are below the one set by Fermi-LAT data from dwarf spheroidal galaxies.

\subsection{Collider phenomenology}
\label{CollPheno}

At the beginning of this section we pointed out that the spectrum of a typical higgsino LSP scenario is characterised by a light higgsino doublet with almost degenerate neutralinos and charginos $\chi_1^0$, $\chi_2^0$ and $\chi_1^\pm$. On the other hand, the rest of the spectrum is heavier and in principle free.\footnote{In the MSSM scalars should be at least at $2$ TeV to have a $125$ GeV Higgs \cite{Vega:2015fna} but in the NMSSM they could be lighter \cite{Hall:2011aa}. On the other hand, in a higgsino LSP scenario binos and winos should satisfy the LUX bounds described above.} In this scenario, the only observable SUSY particle could be a non-thermally produced higgsino which could be as light as $300$ GeV. The collider phenomenology of this scenario would be dominated by hard jet production with large missing energy, i.e. a monojet signal and soft leptons. This signal is produced by a pair of electroweakinos through exchange of $\gamma$, $W^\pm$ or $Z$ gauge bosons in the s-channel together with hard QCD initial state radiation.

Moreover, due to the degeneracy of the charginos $\chi_1^\pm$ with the neutralinos $\chi_1^0$ and $\chi_2^0$, they would probably have a lifetime $\tau\geq0.1$ ns, which is of the order of the collider scale \cite{Low:2014cba}. That makes these charginos long-lived particles which could generate a disappearing track signal.

Ref. \cite{Baer:2014kya} has shown that using monojet and soft leptons, the 3$\sigma$ exclusion limit for the higgsino mass is $250$ GeV with $1000$ fb$^{-1}$ luminosity at $14$ TeV LHC. Given that this bound is less restrictive than the one imposed by Fermi-LAT dSph, the LHC seems to be less interesting for constraining this scenario. Ref. \cite{Low:2014cba} claimed that for a $100$ TeV machine the exclusion could reach higgsinos of $870$ GeV. Moreover, using disappearing tracks in a $100$ TeV collider, it could be possible to exclude higgsinos up to $750$ GeV but also to discover them for masses of almost $600$ GeV. A similar result was found in \cite{Ismail:2016zby} where there is a more systematic study of the uncertainties for a $100$ TeV collider. As can be seen from Fig. \ref{figInd}, this would imply that, unlike the case of thermally produced higgsinos, a future $100$ TeV collider could be able to test completely the scenario of non-thermally produced higgsino LSP. This makes this scenario a very interesting one to be tested in future colliders.

\section{Non-thermal cosmology from string scenarios}
\label{sec:DirectDetection}

As mentioned in Sec. \ref{section1}, the existence of moduli is a generic feature of string theory. These fields parametrise the shape and the size of the extra-dimensions and, at the level of 4D physics, they would mediate fifth forces whose range is inversely proportional to their mass. Given that these new interactions have not been observed, the moduli need to acquire a mass via the process of moduli stabilisation. The mechanism responsible to make the moduli massive fixes also all the main energy scales of a string compactification like the string scale, the Kaluza-Klein scale, the inflationary scale and the SUSY-breaking scale. The presence of such scalar fields has also a very important impact on cosmology since they can both drive inflaton in the very early universe, and affect the post-inflationary evolution of our universe \cite{NTDM,Allahverdi:2013noa,Allahverdi:2014ppa,Aparicio:2015sda,Cicoli:2015bpq}.

In this section we will perform a model dependent analysis of non-thermal higgsino DM production for the two best developed scenarios of moduli stabilisation in type IIB string theory: the Large Volume Scenario \cite{Balasubramanian:2005zx} and the KKLT setup \cite{Kachru:2003aw}. In order to be explicit and set further constraints besides the ones discussed in Sec. \ref{sec:IndirectDetection}, we will consider three different classes of models: LVS with sequestered and non-sequestered SUSY-breaking and KKLT with nilpotent goldstino (see \cite{Aparicio:2015psl} for a detailed discussion of the hierarchy of energy scales for each case).

\subsection{Sequestered LVS models}

A well-studied scenario in type IIB is LVS with the visible sector localised on D3-branes at singularities \cite{Blumenhagen:2009gk,Aparicio:2015psl,Aparicio:2014wxa}. In this model it is possible to achieve a hierarchy between the soft terms and the gravitino mass which is called \textit{sequestering}. The hierarchies are given by:
\be
M_{1/2}\sim \epsilon\ m_\phi \sim \epsilon^2\ m_{3/2} \sim \epsilon^3 M_s \sim \epsilon^4\ M_p\,,
\label{hierc1}
\ee
where $M_{1/2}$ is the gaugino mass, $m_\phi$ is the mass of the lightest modulus, $m_{3/2}$ is the gravitino mass and $M_s$ is the string scale. The hierarchy parameter $\epsilon\ll 1$ can be expressed in terms of the volume of the extra-dimensions $\mc{V}$: $\epsilon\sim 1/\sqrt{\mc{V}}$.\footnote{The volume is measured in units of the string length $\ell_s = M_s^{-1}$.} This framework can allow for two different scenarios depending on whether the soft scalar masses $m_0$ are of the order of the gaugino mass, i.e. $m_0 \simeq M_{1/2}$, or heavier, i.e. $M_{1/2}\simeq \epsilon\ m_0$. The second case corresponds to a split-SUSY like scenario. From \eqref{hierc1} and $T_{\rm r}\simeq \sqrt{m_\phi^3/M_p}$ we find the following relation between $T_{\rm r}$ and gaugino masses:
\be
M_{1/2}\simeq \epsilon \left(T_{\rm r}^2 M_p\right)^{1/3}\,.
\label{DirectRel}
\ee

\begin{figure}[!t]
\centering
\hspace{-0.5cm}
\includegraphics[width=8.4cm]{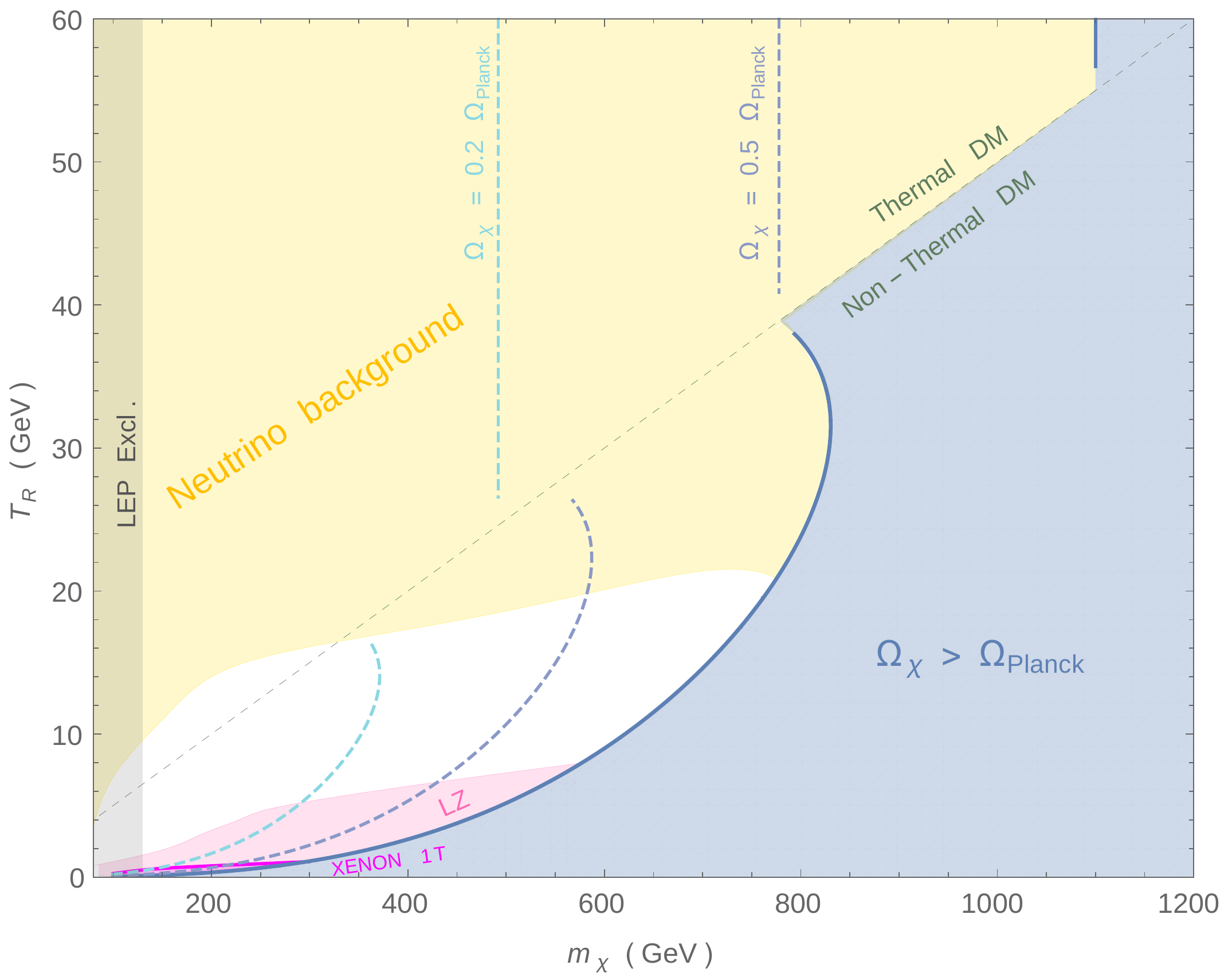}
\hspace{-0.1cm}
\includegraphics[width=8.4cm]{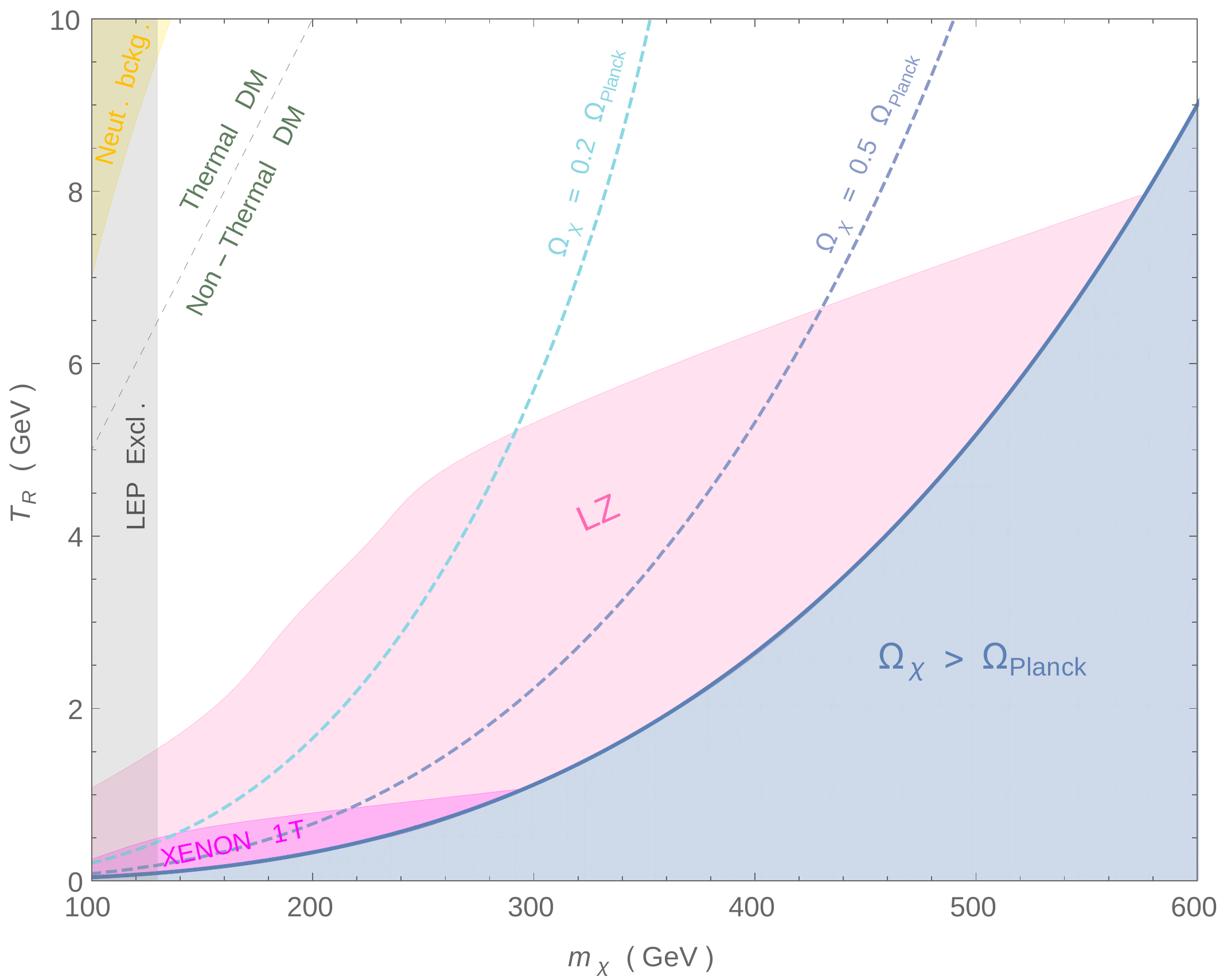}
\hspace{-0.5cm}
\caption{Reheating temperature and DM relic density in terms of the higgsino mass. The different coloured regions indicate bounds from direct detection experiments for $\mu>0$. The yellow coloured region corresponds to the neutrino background. The two pink regions show the sensitivity to direct detection experiments, in particular to XENON1T and LZ. The plot on the right is a zoom on the region with $T_{\rm r}\leq 10$ GeV.}
\label{Fig2}
\end{figure}

\begin{figure}[!t]
\centering
\hspace{-0.5cm}
\includegraphics[width=8.4cm]{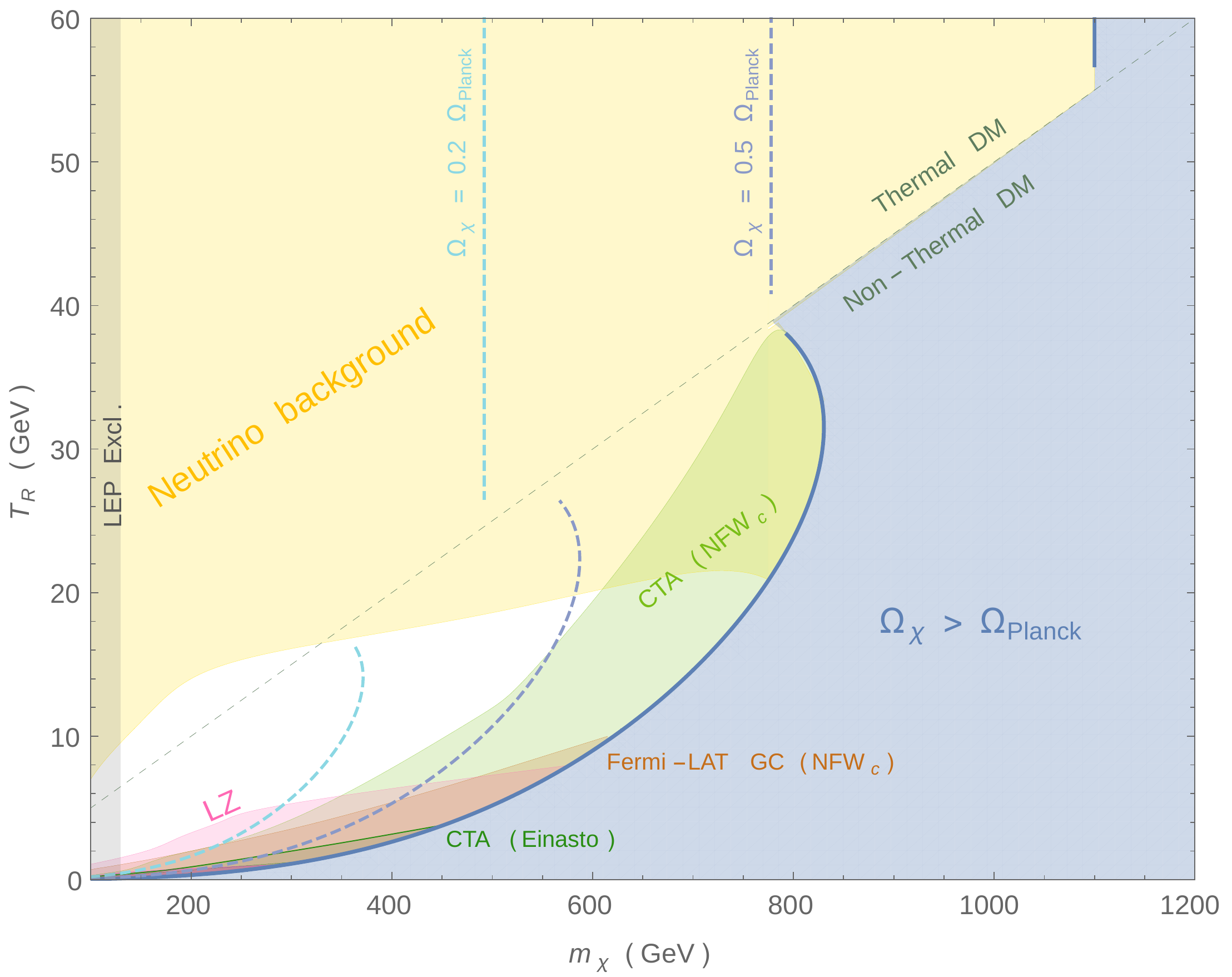}
\hspace{-0.1cm}
\includegraphics[width=8.4cm]{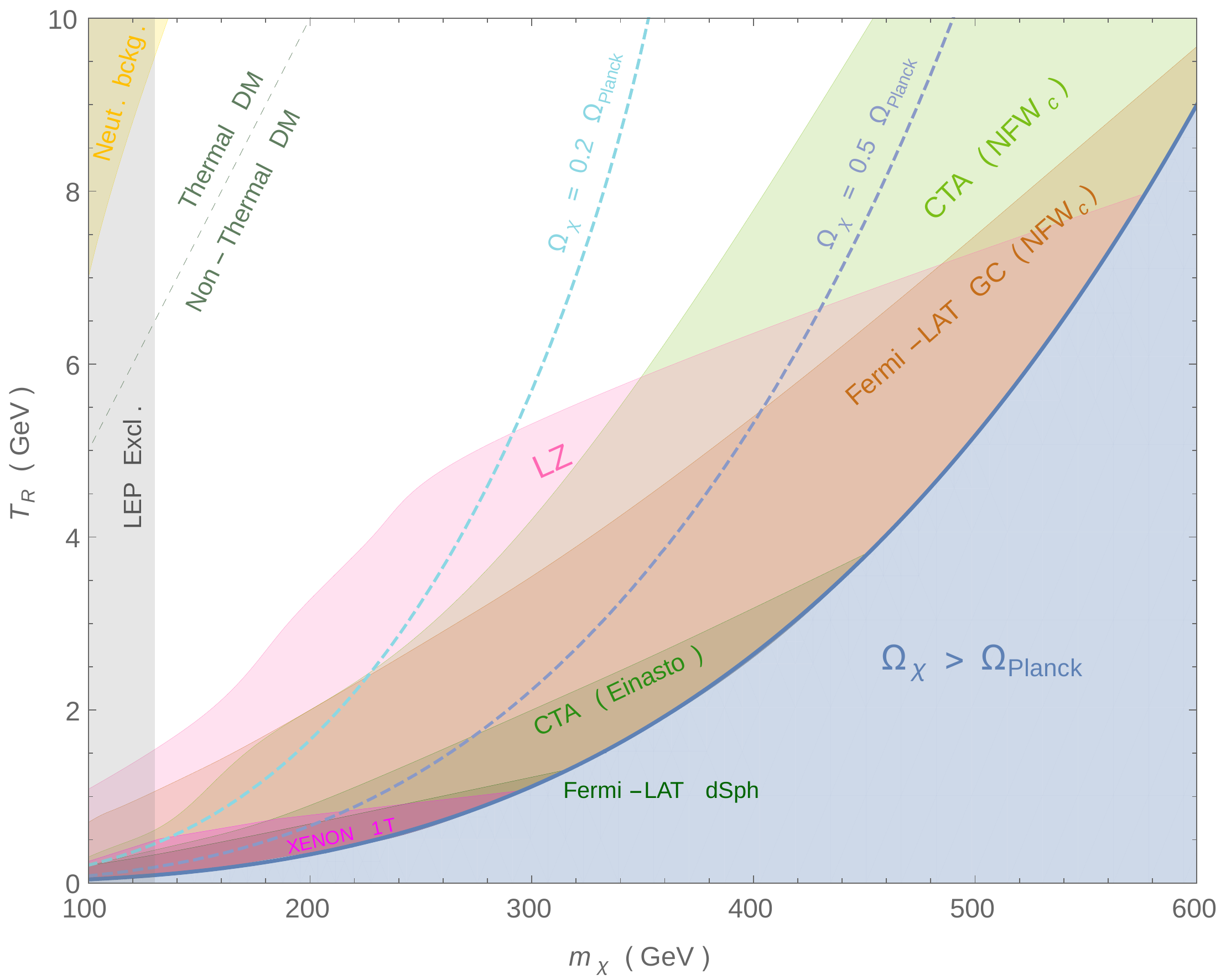}
\hspace{-0.5cm}
\caption{Combined DM indirect and direct detection bounds from Fig. \ref{figInd} and \ref{Fig2}. The plot on the right is a zoom on the region with $T_{\rm r}\leq10$ GeV.}
\label{Fig3}
\end{figure}

\begin{figure}[!t]
\centering
\hspace{-0.5cm}
\includegraphics[width=8.4cm]{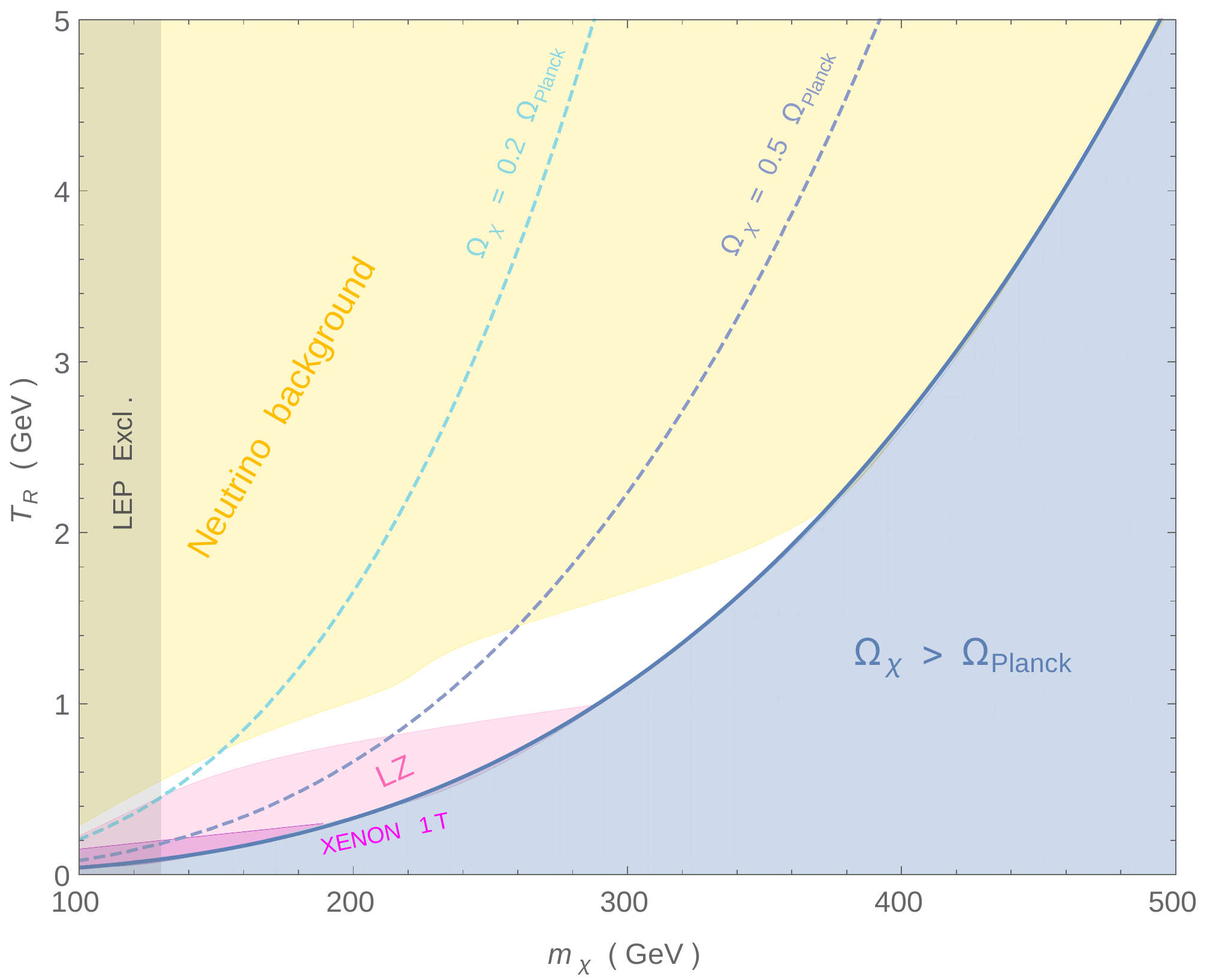}
\hspace{-0.1cm}
\includegraphics[width=8.4cm]{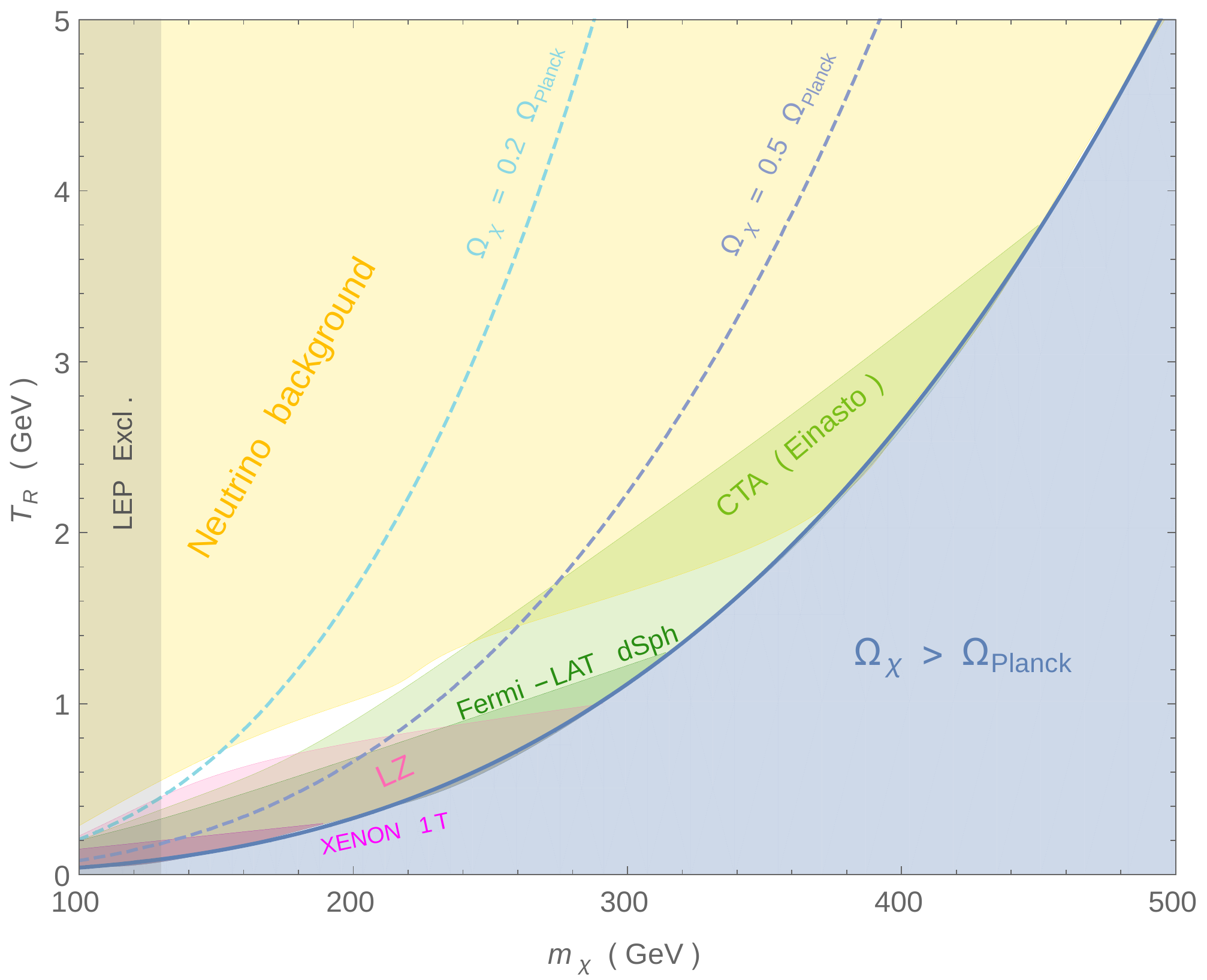}
\hspace{-0.5cm}
\caption{Negative $\mu$ case analysis of DM direct detection bounds. The plot on the right shows the combined DM indirect and direct detection bounds.}
\label{Fig4}
\end{figure}

Let us now consider sequestered LVS models with non-thermal higgsino DM production described in Sec. \ref{sec:IndirectDetection}. For a given value of $\epsilon$ (or equivalently for a fixed value of $\mc{V}$), (\ref{DirectRel}) gives $T_{\rm r}$ in terms of $M_{1/2}$. Substituting this relation in \eqref{NTann} we find that the non-thermal DM relic density depends on the ratio between higgsino and gaugino masses. The hierarchy $M_{1/2}-\mu$ is interesting because it allows us to introduce DM direct detection bounds. 

We consider a particularly interesting value of the extra-dimensional volume, $\V \simeq 10^7$, because it yields both a string scale which is high enough to allow for GUT theories and viable inflationary model building, $M_s \sim M_{\rm GUT} \sim 10^{16}$ GeV, and low-energy gaugino masses around $\mc{O}(1-10)$ TeV. Using \texttt{micrOMEGAs} we have computed the spin independent (SI) cross section and compared it with prospects from XENON1T \cite{Aprile:2011zza} and LZ \cite{Malling:2011va} (the bounds from LUX are irrelevant in this scenario). The relation \eqref{DirectRel} allows us to project all this information in the $(T_{\rm r}, m_\chi)$-plane which is the same parameter space used in Fig. \ref{figInd}.

In Fig. \ref{Fig2} we show the impact of direct detection bounds on the underlying parameter space for the case with $\mu > 0$. We see that the sensitivity to direct detection is generically small. Large scale DM direct detectors (beyond $1$ Ton) are necessary to cover the region with reheating temperatures close to $10$ GeV. In Fig. \ref{Fig3} we show a comparison between direct and indirect detection sensitivity. One can see that Fermi-LAT is already restricting more than what XENON1T can do. In order to constrain the parameter space more than what DM indirect detection is already doing, it is therefore necessary to consider experiments like LZ. Finally, for $T_{\rm r}>20$ GeV, which corresponds to moduli masses around $m_\phi\gtrsim \mc{O}(10^4)$ TeV, the neutrino background covers the entire remaining parameter space.

In Fig. \ref{Fig4} we show the same analysis for the case with $\mu<0$. Note that the sensitivity to DM direct detection bounds is much lower than for the positive $\mu$ case. The reason can be understood from the effective $h\chi\chi$ coupling which appears in the nucleon-neutralino interaction:
\be
c_{h\chi\chi}\sim \frac{1+\textrm{sign}(\mu)\sin2\beta}{M_1-\textrm{sign}(\mu)\mu}\,,
\ee
where for $\mu<0$ the SI cross section tends to be smaller. Unlike the scenario with $\mu>0$, even large scale detectors like LZ will induce constraints below the Fermi-LAT dSph bounds.

The neutrino background seems to be larger than the signal for most regions of the parameter space. This means that DM direct detection experiments will hardly be able to probe this region. In \cite{Dent:2016iht} the authors have investigated the recoil spectra from different DM-nucleon effective field theory operators and they have compared them to the nuclear recoil energy spectra that are predicted to be induced by astrophysical neutrino sources. The dominant MSSM SI neutralino-nucleon operators ($\bar{q}q\bar{\chi}\chi$) can be distinguished from the neutrino backgrounds for a very large exposure, $10^3$ tonne years, since the recoil spectra for the signal is similar to the background. 

From Figs. \ref{Fig2} - \ref{Fig4} we can extract another interesting information about the sparticle spectrum. Given that in the sequestered LVS scenario $M_{1/2}$ is universal at the GUT scale, binos, winos and gluinos have different masses. If the DM relic density is saturated by higgsinos with $m_\chi\simeq 300$ GeV, we have $m_{\tilde{B}}\simeq 1.9$ TeV, $m_{\tilde{W}}\simeq 3.8$ TeV and $m_{\tilde{g}}\simeq 10.2$ TeV. Another interesting situation would be the case with $m_\chi\simeq 600$ GeV since it is in the region close to the LZ detection reach. In this case the spectrum of gauginos would be $m_{\tilde{B}}\simeq 6$ TeV, $m_{\tilde{W}}\simeq 12.3$ TeV and $m_{\tilde{g}}\simeq 33.3$ TeV. In both cases sfermion masses are at least on the multi-TeV range (typically $\mc{O}(10)$ TeV) or heavier (their detailed spectrum depends on whether the SUSY model is split-like or not).

Finally, it is worth commenting that it is not clear whether the GUT boundary conditions of the sequestered LVS scenario allow for a light higgsino LSP. For example, a split-SUSY case with universal scalar masses (see \cite{Aparicio:2015psl}) would not allow light higgsinos. In this scenario the higgsino would actually be so heavy to induce a large loop correction to both the wino and bino masses, making them heavy as well. The result is a gluino LSP scenario which is already ruled out. In the case where $m_0\sim M_{1/2}$, the determination of the GUT boundary conditions which allow for light higgsinos is still an open question. It seems to be a set of very special conditions which allow for a focus-point behaviour (see for instance \cite{Aparicio:2015sda}). Hence the LVS sequestered scenario requires further studies to check if it has enough freedom to realise the higgsino LSP case studied in Sec. \ref{sec:IndirectDetection}.

\subsection{Non-sequestered LVS models}
\label{sec:NonSequestered}

An alternative option for the realisation of the visible sector is to localise SM gauge interactions on stacks of D7-branes wrapping some sub-manifolds of the compact space. In this case the gauge degrees of freedom are directly coupled to the sources of SUSY-breaking, and so all soft terms are of the same order of the gravitino mass but heavier than the lightest modulus $\phi$ \cite{Aparicio:2015psl}: 
\be
m_\phi \sim \epsilon \left ( M_{1/2}\sim M_0 \sim m_{3/2} \right)  \sim \epsilon^2 M_s \sim  \epsilon^3 M_p\,.
\label{hierc2}
\ee
In order to avoid the cosmological moduli problem we require $T_{\rm r}\geq 4$ MeV \cite{Hannestad:2004px}, and hence the modulus mass becomes $m_\phi\geq 34$ TeV. In turn, \eqref{hierc2} implies that all soft terms are very heavy, $M_{1/2}\gg 1 $ TeV. In particular, such heavy gauginos induce a large one-loop contribution to the higgsino mass \cite{Hall:2011jd}:
\be
\Delta m_{\chi} = -\frac{\sin(2\beta)}{32 \pi^2}\left[ 3g_2^2 M_2 \log\left(\frac{M_H}{M_2}\right) + g_1^2 M_1 \log\left(\frac{M_H}{M_1}\right) \right]\,,
\label{loop}
\ee
where $M_H$ is the mass of the heavy Higgs in SUSY models. 

\begin{figure}[!t]
\centering
\includegraphics[scale=0.5]{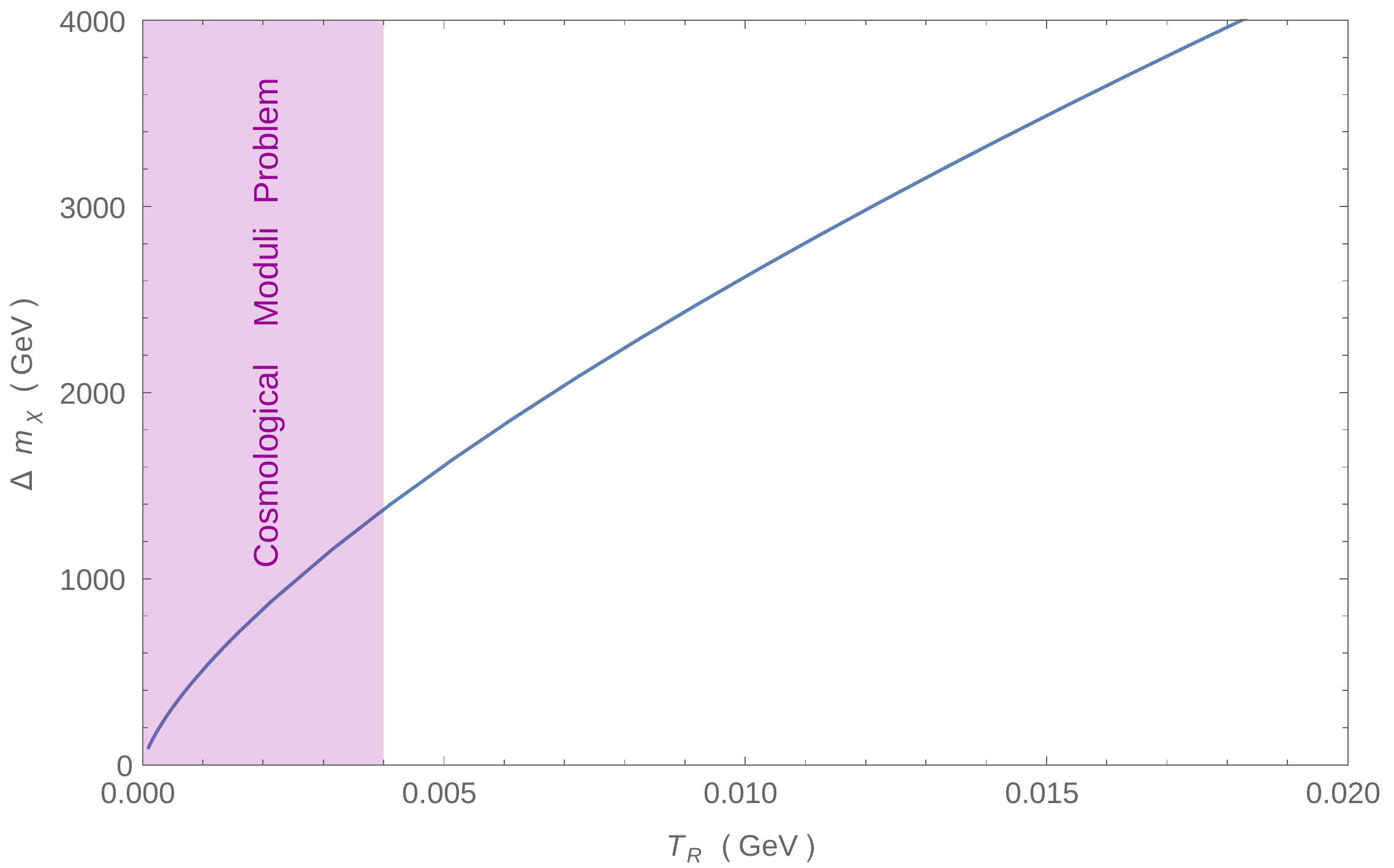}
\caption{One-loop contribution to higgsino mass induced by heavy gauginos as a function of the reheating temperature $T_{\rm r}$ due to the modulus decay.}
\label{Fig5}
\end{figure}

If this one-loop induced mass is very large, it could make the higgsino dangerously heavy. Fig. \ref{Fig5} shows this contribution in terms of $T_{\rm r}$. We have performed this computation using the hierarchies in \eqref{hierc2} and expressing them in terms of the reheating temperature (assuming again that $\mc{V}\sim 10^7$). The contribution of $\tan\beta$ to \eqref{loop} has been calculated recursively in order to obtain a Higgs mass of $125$ GeV by using \texttt{SUSYHD} \cite{Vega:2015fna}. For temperatures above $4$ MeV, the one-loop induced higgsino mass becomes $\Delta m_\chi\geq1.4$ TeV. As can be seen from Fig. \ref{figInd}, this value of the higgsino mass leads to DM overproduction. Hence non-sequestered LVS models needs R-parity violation in order to avoid the overclosure of the universe. It would then be necessary to look for both alternative DM explanations and a mechanism to avoid fast proton decay in GUT theories.

\subsection{KKLT models with nilpotent goldstino}

Moduli stabilisation for KKLT models with a dS vacuum generated by anti D3-branes has been recently discussed in \cite{Aparicio:2015psl}. In this scenario the hierarchy between the gravitino and the lightest modulus mass is:
\be
m_{3/2} \sim \epsilon^{4/3}\ m_{\phi} \sim \epsilon^2 M_p\,.
\label{hierc3}
\ee
It is easy to see from \eqref{hierc3} that the gravitino is lighter than the modulus. Hence in KKLT models the last decaying particle which dominates the thermodynamic history of the universe is not a modulus but the gravitino. The gravitino is coupled to other particles only gravitationally, so DM production can be described using the same techniques illustrated in Sec. \ref{section2}. However, there is a difference with respect to the modulus case: gravitinos are not originated by a misalignment mechanism but rather by the inflaton decay. Hence this scenario is more model dependent because it depends on the scale of inflation. For instance, if inflation ends so late that the inflaton cannot kinematically decay into gravitinos, those will not be produced unless the last decaying modulus would be able to produce them.

The hierarchy between gravitino and scalar and gaugino masses is instead given by \cite{Aparicio:2015psl}:
\begin{eqnarray}
\label{kkltScal}
m	_0 &\sim & m_{3/2} \ \ \ \ \textrm{for visible sector on D7-branes,}\\
m_0 &\sim &\epsilon\ m_{3/2} \ \ \textrm{for visible sector on D3-branes,}
\end{eqnarray}
\be
M_a\simeq \left(\frac{\alpha_a b_a}{4\pi} + \epsilon^{4/3}\right) m_{3/2}\,,
\label{kkltGaug}
\ee
where the first term in \eqref{kkltGaug} is the anomaly mediation contribution. A big difference with respect to the LVS scenario is that in KKLT the anomaly mediation contribution to gaugino masses dominates over the moduli mediation one. From \eqref{kkltScal} - \eqref{kkltGaug} it can be seen that there are two KKLT scenarios: typical anomaly mediation mini split-SUSY models when the visible sector is on D7-branes, and SUSY models with anomaly mediated gauginos which are a bit lighter than sfermions for the visible sector on D3-branes. 

\begin{figure}[!t]
\centering
\includegraphics[scale=0.45]{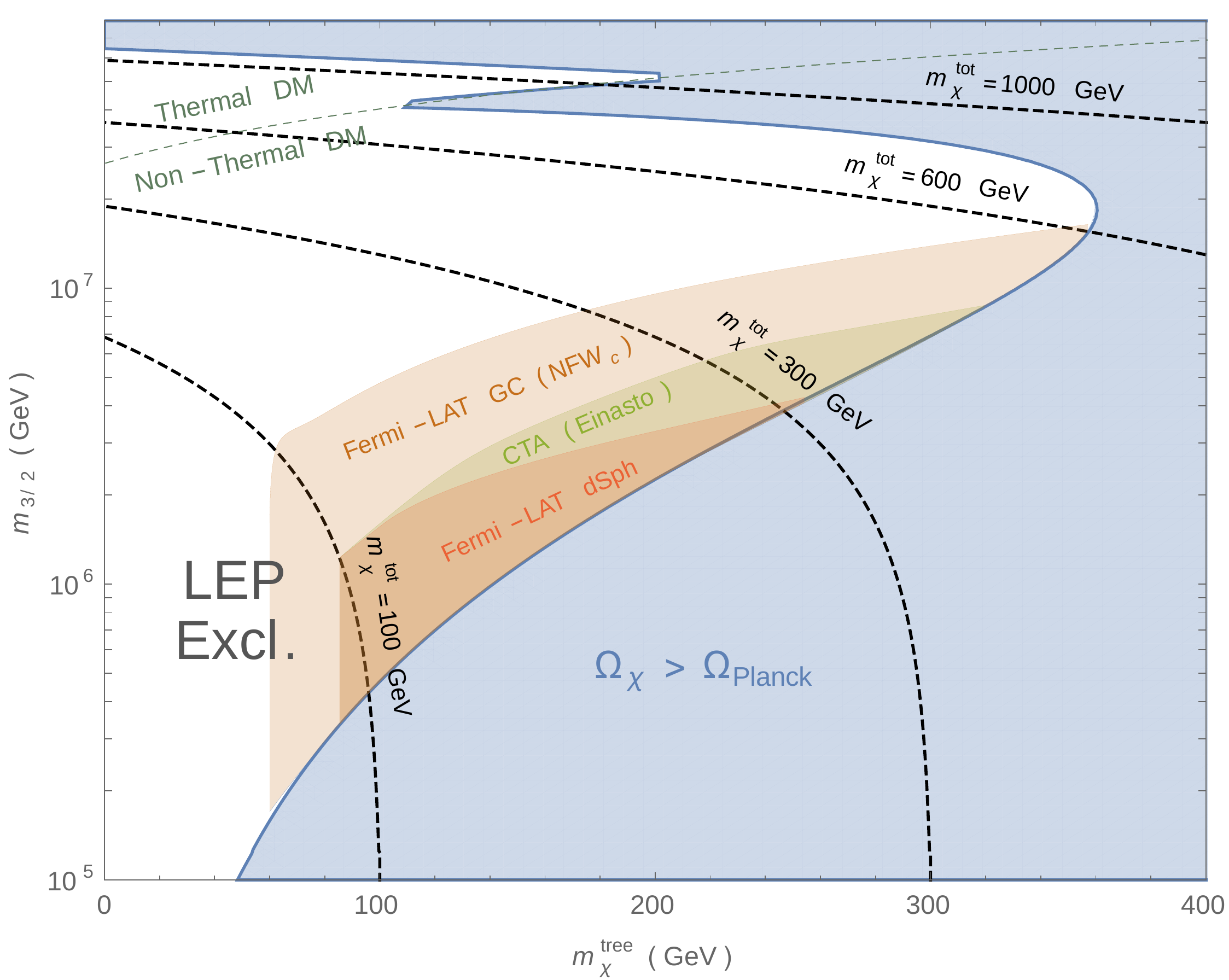}
\caption{Gravitino mass and DM relic density in terms of higgsino mass. The different coloured regions indicate bounds from indirect detection experiments. The red and orange ones correspond to the actual Fermi-LAT pass 8 bounds. The green line corresponds to CTA prospects. The diagonal dashed green line refers to $T_{3/2} = T_{\rm r} = m_\chi/20$.}
\label{FigKKLT}
\end{figure}

Like in the LVS case, in order to preserve BBN results we impose $T_{3/2} > 4$ MeV which implies $m_{3/2} \geq 10^5$ GeV \cite{Nakamura:2006uc}. From \eqref{kkltScal} - \eqref{kkltGaug} we can immediately see that this lower limit on the gravitino mass pushes scalars and gauginos to heavy scales. This has important consequences on the higgsino LSP scenario. In particular, similarly to the non-sequestered LVS case, heavy gauginos induce large one-loop contributions to the higgsino mass which tend to push higgsinos to heavy scales where their abundance would overclose the universe. 

However, if there is a leading order cancellation between the two contributions to gaugino masses in \eqref{kkltGaug}, the hierarchy between gauginos and gravitinos could be larger. This could allow for a region where the higgsino is still a good DM candidate. After studying this situation, we have found that in KKLT models with the visible sector on D7-branes higgsino are always too heavy. 

On the other hand, if the visible sector lives on D3-branes, there is a region where higgsino DM is still possible. This is due to the combination of the small hierarchy between scalars and gravitinos from \eqref{kkltScal} and the large hierarchy between gauginos and gravitinos which can be arranged by tuning  the two different contributions in \eqref{kkltGaug}.\footnote{Technically this region can be obtained by setting the non-perturbative effect number $N=4$ and $\mc{V}\simeq 10^3$. In KKLT the internal volume is bounded both from below in order to trust the effective field theory, $\mc{V}\geq 10^3$, and from above  to avoid tachyonic sleptons from anomaly mediation, $\mc{V}\leq 10^5$ \cite{Aparicio:2015psl}.}

The results of this analysis are presented in Fig. \ref{FigKKLT} which shows that the one-loop contribution to the higgsino mass from heavy gauginos (see \eqref{loop}) sets an upper bound on the gravitino mass, $m_{3/2}\lesssim 6\times 10^7$, (the lower bound comes from BBN) beyond which there is DM overproduction. The dashed black lines show the total higgsino mass compared with the tree level one (corresponding to $\mu$) plotted on the x-axis. It is interesting to notice that even for $\mu = 0$ there could be a thermal higgsino LSP of $1.1$ TeV generated completely at loop level for $m_{3/2} \simeq 6\times 10^4$ TeV. This would be reproduce a spread SUSY scenario with higgsino LSP \cite{Hall:2011jd}. Note also that spread SUSY cannot be realised for the non-thermal case since it requires $\mu \neq 0$.

Moreover, Fig. \ref{FigKKLT} illustrates very clearly the effect of a late decaying gravitino on the DM abundance. The green dashed line corresponds to $T_{3/2} = T_{\rm f} \simeq m_\chi^{\rm tot}/20$ and separates the region where the gravitino does not affect DM production since it decays before the thermal freeze-out of the higgsino LSP, from the region with $T_{3/2} < T_{\rm f}$ where the gravitino decay has the same effects as those described in Sec. \ref{sec:IndirectDetection} for the modulus decay. Therefore the results shown in Fig. \ref{FigKKLT} are the same as those of Fig. \ref{figInd} with the only difference being that they are plotted in terms of $m_{3/2}$ instead of $T_{\rm r}$. 

Finally Fig. \ref{FigKKLT} indicates that non-thermally produced higgsinos with $m_\chi \simeq 300$ GeV require a gravitino mass of order $m_{3/2}\simeq 4\times 10^3$ TeV. This, in turn, gives scalars around $m_0\simeq 100$ TeV and gaugino masses of order $m_{\tilde{B}}\simeq 11$ TeV, $m_{\tilde{W}}\simeq 22$ TeV and $m_{\tilde{g}}\simeq 60$ TeV. This implies that higgsinos of $300$ GeV are in a region where the higgsino-nucleon SI cross section is almost below the neutrino background for $\mu>0$ and completely inside the neutrino background for $\mu<0$. Therefore it seems that DM direct detection is much less useful in the KKLT scenario than in the LVS one.

\section{Conclusions}
\label{sec:Conclusions}

In this work we focused on supersymmetric models where the LSP is a higgsino-like neutralino which plays the role of DM in the context of a non-standard cosmology. The difference with respect to the standard cosmological history comes from the presence of new degrees of freedom which can decay late changing the DM relic abundance produced by the standard thermal freeze-out scenario. The presence of such fields is well motivated from string theory where moduli fields naturally emerge in its low-energy 4D limit. 

The paper is divided into two parts. In Sec. \ref{section1}, \ref{section2} and \ref{sec:IndirectDetection} we performed a model independent analysis of supersymmetric models with non-thermal production of light higgsino DM. In Sec. \ref{sec:DirectDetection} we presented instead a model dependent discussion of different string models where a non-standard cosmology is motivated by the presence of moduli which decay at late time. For each string model we studied theoretical and observational constraints on higgsino  non-thermal DM production.

The main conclusions of the model independent analysis developed in the first part of the paper are:
\begin{enumerate}
\item In non-thermal cosmologies with an extra period of matter domination which ends via reheating with temperatures of $\mc{O}(1 - 10)$ GeV (above BBN), light higgsinos with masses as low as a few hundred GeV can correctly saturate the DM content measured by the Planck satellite. 

\item Such light higgsinos are very interesting from both a theoretical and an experimental point of view. The fact that they are very light makes them easily accessible to both indirect detection and collider searches.

\item The strongest bound from indirect detection imposes that non-thermally produced higgsinos cannot be lighter than $300$ GeV. This bound comes from Fermi-LAT dSph where the dependence on the DM astrophysical profile is less important than in galactic centre observations. Observations by future experiments like CTA, together with data from Fermi-LAT GC, could cover essentially the entire parameter space of this scenario. On the other hand, unlike in the thermal case, collider signals from a $100$ TeV machine could test directly this scenario using searches on monojet and disappearing tracks.
\end{enumerate}

From the model dependent discussion performed in the second part of the paper, we can conclude that:
\begin{enumerate}
\item The main difference between LVS and KKLT scenarios for type IIB moduli stabilisation is that the last particle to decay in LVS models is the lightest modulus, while in KKLT models it is the gravitino. However, both cases feature a late decaying particle (scalar in LVS and fermion in KKLT) which motivates the analysis performed in the first part of the paper. Depending on the scenario under consideration, the hierarchy between the masses of the moduli, the higgsinos and the other superpartners can take a different form. 

\item When the visible sector is localised on D7-branes, both LVS and KKLT models with stable higgsino LSP are plagued by the problem of DM overproduction since heavy gaugino masses give rise a large contribution to higgsino masses at one-loop level.

\item KKLT models with the visible sector on D3-branes still tend to have problems with higgsino DM overproduction due to the fact that gauginos are heavy in order to have gravitinos which decay before BBN. However, there is a fine-tuned region of the underlying parameter space where the non-thermal production of light higgsinos can yield the correct DM abundance. 

\item LVS models with the visible sector on D3-branes seem to be the best option to realise non-thermal scenarios with light higgsino DM. In fact, one-loop corrections to higgsino masses are small since sequestering effects suppress gaugino masses with respect to the mass of the decaying modulus. By exploiting the relation between the modulus and the gaugino mass, we managed to rewrite the reheating temperature in terms of the gaugino mass. This allowed us to introduce the effect of DM direct detection searches. We have found that, on the one hand, it is necessary to use large scale DM direct detection experiments (beyond $1$ Ton) to constrain more than what indirect detection already does, while, on the other hand, a large region of the parameter space falls below the neutrino background, and so DM direct detection experiments do not seem to be very useful to explore the parameter space of these theories.
\end{enumerate}

Future experiments will be able to completely probe the underlying parameter space of supersymmetric models with non-thermal light higgsino DM. This makes this scenario very interesting from both DM detection and future collider searches at $100$ TeV and motivates a detailed analysis from both sides.

\section*{Acknowledgments}

We would like to thank X. Chu, M. Goodsell, G. Grilli di Cortona, E. Hardy, S. Krippendorf, G. Villadoro, C. Yaguna and G. Zaharijas for useful conversations. The work of B.D. is supported in part by DOE Grant DE-FG02-13ER42020.

\end{document}